\documentclass[conference]{IEEEtran}
\IEEEoverridecommandlockouts
\usepackage{amsmath, amssymb, bm, cite, epsfig, psfrag, mathtools}
\usepackage{xcolor}
\usepackage{array}
\usepackage{algorithm}
\usepackage{algpseudocode}

\def\BibTeX{{\rm B\kern-.05em{\sc i\kern-.025em b}\kern-.08em
		T\kern-.1667em\lower.7ex\hbox{E}\kern-.125emX}}

\begin{document}
\makeatletter
\makeatother
\title{Can Massive MIMO Support URLLC?}

\author{\IEEEauthorblockN{Hangsong Yan}
	\IEEEauthorblockA{\textit{NYU Wireless} \\
		New York, US\\
		hy942@nyu.edu}
	\and
	\IEEEauthorblockN{Alexei Ashikhmin}
	\IEEEauthorblockA{\textit{Bell Labs, Nokia} \\
		New Jersey, US \\
		alexei.ashikhmin@nokia-bell-labs.com}
	\and
	\IEEEauthorblockN{Hong Yang}
	\IEEEauthorblockA{\textit{Bell Labs, Nokia} \\
		New Jersey, US \\
		h.yang@nokia-bell-labs.com}}
	

\maketitle

\begin{abstract}
	We investigate the feasibility of using Massive MIMO to support URLLC in both coherence interval based and 3GPP compliant pilot settings. We consider grant-free uplink transmission with MMSE receiver and adopt 3GPP channel models. In the coherence interval based pilot setting, by extensive system level simulations, we find that using a Massive MIMO base station with 128 antennas and MMSE receiver, URLLC requirements can be achieved in Urban Macro (UMa) Non-Line of Sight (NLoS) with orthogonal pilots and Neyman-Pearson detector. However, in the 3GPP compliant pilot setting, even by using the covariance matrix of Physical Resource Block (PRB) subcarriers for active UE detection and channel estimation as well as open-loop power control, we find that URLLC requirements are still challenging to achieve due to the insufficient pilot length and pilot symbol location regulations in a PRB.
\end{abstract}


\section{Introduction}\label{Sec:Introduction}
ITU defined three main categories for 5G communications -- eMBB (enhanced Mobile BroadBand), URLLC (Ultra-Reliable Low-Latency Communication), and mMTC (massive Machine Type Communications) \cite{ITU_M2410}. URLLC is required by Industry 4.0 applications such as robot motion control and process automation remote control. Its use cases can be established in many industrial sectors including manufacturing, healthcare, transportation, new energy exploration, and entertainment. URLLC specifications include ultra-high reliability (e.g., $99.999$\%, sometimes also $99.9999$\%~\cite{3GPP_38_824}) and ultra-low latency (e.g., 1 ms). At the same time, relatively high data throughput is also required (e.g., 250 bytes per 1 ms~\cite{3GPP_38_824}). 
To tackle these challenging requirements, the industry standards have proposed to use grant free (GF) access protocol \cite{3GPP_38_214} to reduce scheduling latency. However, this results in multi-UE collision thus requires accurate active UE detection.

We consider the feasibility of using Massive MIMO (mMIMO) technology to support URLLC in two pilot settings. 
In the coherence interval based pilot setting, both long orthogonal pilots and short non-orthogonal Gold sequence are considered. By incorporating active UE detection, LMMSE channel estimation, and MMSE MIMO receiver, extensive simulations are conducted to quantify the uplink performance of a URLLC system supported by a single mMIMO base station (BS) in both i.i.d. Rayleigh fading and 3GPP channel models~\cite{3GPP_38_901}. For the 3GPP compliant pilot setting, to tackle with the pilot length limit and the channel coefficients variation across subcarriers in a physical resource block (PRB), we incorporate the channel coefficients covariance matrix across different subcarriers in a PRB into the active UE detector and LMMSE channel estimator. By applying a 3GPP standards compliant open-loop power control, per-UE effective throughput performance is evaluated numerically. 
\vspace{-2.1mm}

\section{System Model}\label{System Model}
We consider a single cell mMIMO URLLC network. One BS with $M$ antennas is located at the center of the cell. In total $\tilde{K}$ single antenna UEs are uniformly distributed in the whole cell. 
We consider GF access in our work. At the beginning of a certain time duration, $K$ out of $\tilde{K}$ UEs are authorized to have GF access for URLLC service and each of these $K$ UEs is assigned a unique pilot. At a given moment of this time duration, $K_a$ of $K$ UEs are active, where $K_a$ is modeled by a Poisson distribution. We consider orthogonal frequency division multiplexing (OFDM) in our system and the coherence interval length is denoted as $\tau_c$ in terms of OFDM symbols. We consider two pilot settings. In the first setting, pilots occupy $\tau$ of $\tau_c$ OFDM symbols for every subcarrier and the channel coefficient of each subcarrier is constant in a coherence interval and independent in different coherence intervals. We call this setting as coherence interval based pilot setting. In the second 3GPP compliant pilot setting, pilot symbols are distributed in a PRB with an example illustrated in Fig.~\ref{fig:DM_RS_Locations} and channel estimation is performed per PRB. Note that Gold sequence is applied in both settings, but the distribution of pilot symbols in the time-frequency resource is different. Also note that we focus on physical layer analysis on mMIMO supported URLLC in this paper, an analysis on balancing queueing and retransmission can be found in~\cite{Du_Balanc_2020}.

We denote by $g_{m,k}^{n}$ the channel coefficient between the $m$th antenna of the BS and the $k$th UE at the first subcarrier of the $n$th subband. It is modeled by
\begin{equation}
	g_{m,k}^{n} = \sqrt{\beta_k}h_{m,k}^n,
\end{equation}
where $\beta_{k}$ is the large-scale fading coefficient of the $k$th UE and $h_{m,k}^n$ is the small-scale fading coefficient which is antenna and frequency dependent. For a given bandwidth (BW), there are $N$ independent subbands and the BW of each subband depends on the coherence BW.

\section{Coherence Interval Based Pilots}
\subsection{Received Signal for UE Detection and Channel Estimation}
The received signal at the $m$th antenna of the BS and the $n$th subband is expressed as
\vspace{2mm}
\begin{equation}
	\label{eq:AUD_received_signal}
	\mathbf{y}_m^n = \boldsymbol{\Phi}\mathbf{A} \bar{\mathbf{g}}_{[m]}^n + \mathbf{w}_m^n,
\end{equation}
where $\boldsymbol{\Phi} = [\boldsymbol{\phi}_1\, \boldsymbol{\phi}_2\, ...\, \boldsymbol{\phi}_K] \in \mathbb{C}^{\tau \times K}$ is the pilot matrix with $\tau$ denotes the pilot length
and $\|\boldsymbol{\phi}_k\|_2 = 1$ for $k = 1, 2, ..., K$. $\mathbf{A}$ is the indicator matrix defined as $\mathbf{A} = \text{diag}\{[a_1,a_2,...,a_K]\}$ with $a_k = 1$ if the $k$th UE is active and $a_k = 0$ if the $k$th UE is not active. 
$\bar{\mathbf{g}}_{[m]}^n = \sqrt{\tau \rho_p} [g_{m, 1}^n, g_{m, 2}^n, ..., g_{m, K}^n]^T$
where $\rho_{p}$ is the normalized signal-to-noise ratio (SNR) of each pilot symbol. 
$\mathbf{w}_m^n$ is the noise vector with $\sim i.i.d.\ \mathcal{CN}(0, 1)$ entries. 

\subsection{Active UE Detection with Orthogonal Pilots}\label{Sec:Active UE Detection with Orthogonal Pilots}
To use orthogonal pilots, we must have $\tau \ge K$. We assume $\tau = K$ to minimize the pilot overhead. Then $\boldsymbol{\phi}_k^H \boldsymbol{\phi}_{k'}= \delta(k - k' )$ for $k, k' = 1, 2, ..., K$ and the optimal Neyman-Pearson (NP) \cite{Kay_1993_Detection} detector can be used for active UE detection.
We denote $\mathbf{f}_{k}$ as the pre-processing operator for the $k$th UE detection. Due to the application of orthogonal pilots, the signal used for the $k$th UE detection, $z_{m,k}^n$, can be obtained by a simple correlation operation on $\mathbf{y}_m^n$:
\begin{equation}
	\label{eq: z}
	z_{m,k}^n=\boldsymbol{f}_k^H \mathbf{y}_m^n = \sqrt{\tau \rho_p}\sum_{k' \in \mathcal{A}} \boldsymbol{\phi}_k^H \boldsymbol{\phi}_{k'} g_{m,k'}^n + \boldsymbol{\phi}_k^H \mathbf{w}_m^n,
\end{equation}
where set $\mathcal{A}$ is the active UE set. Note that if the $k$th UE is active, $z_{m,k}^n = g_{m,k}^n + \boldsymbol{\phi}_k^H \mathbf{w}_m^n \sim \mathcal{CN}(0, \tau\rho_p\beta_k + 1)$ and if the $k$th UE is not active, $z_{m,k}^n = \boldsymbol{\phi}_k^H \mathbf{w}_m^n \sim \mathcal{CN}(0, 1)$. 
Define 
$\mathbf{z}_k = [\hat{\mathbf{z}}_{1,k}^1, \hat{\mathbf{z}}_{2,k}^1,..., \hat{\mathbf{z}}_{M,k}^1, \hat{\mathbf{z}}_{1,k}^2,...,\hat{\mathbf{z}}_{m,k}^n,...,\hat{\mathbf{z}}_{M,k}^N]^T \in \mathbb{C}^{MN \times 1}$ where $\hat{\mathbf{z}}_{m,k}^n=[Re\{z_{m,k}^n\}, Im\{z_{m,k}^n\}]$. 
Based on the $i.i.d.\ \mathcal{CN}$ assumption of $\{h_{m,k}^n\}$, we can obtain the following two hypotheses about whether the $k$th UE is active:
\begin{equation}
\begin{aligned}
\label{eq:hypotheses}
	&\mathcal{H}_0 (\text{i.e., inactive}): \mathbf{z}_k \sim \mathcal{N}(\mathbf{0},(\sigma_{0, k}^2/2) \mathbf{I}_{2MN}),\\
	&\mathcal{H}_1 (\text{i.e., active}): \mathbf{z}_k \sim \mathcal{N}(\mathbf{0}, (\sigma_{1, k}^2/2) \mathbf{I}_{2MN}),
\end{aligned}
\end{equation}
where $\sigma_{0, k}^2 = 1$ and  $\sigma_{1, k}^2 = \tau\rho_p\beta_k + 1$. 
We define the test statistics as $T(\mathbf{z}_k) = \sum_{n=1}^{N}\sum_{m=1}^{M}|z_{m,k}^n|^2$. According to (\ref{eq:hypotheses}) and the NP detector procedures given in \cite{Kay_1993_Detection}, the $k$th UE is identified as active if
\begin{equation}
	\label{eq:T_z_k}
	T(\mathbf{z}_k) > \gamma',\ 
	\gamma' = Q_{\mathcal{X}_{2MN}^2}^{-1}(\text{P}_\text{FA})\sigma_{0, k}^2/2,
\end{equation}
where $Q_{\mathcal{X}_{2MN}^2}(\cdot)$ is the complementary cumulative distribution function of $\mathcal{X}_{2MN}^2$, which is the Chi-square distribution with $2MN$ degree of freedom and $\text{P}_\text{FA}$ is a predetermined probability of false alarm.


\subsection{Active UE Detection with Gold Sequence}
The application of short non-orthogonal pilots as configured in the 3GPP standards (e.g., Gold sequence) means $\tau < K$. Due to this fact, the approach of simple correlation pre-processing + NP detector does not work properly for Gold sequence. 
In this section, we adopt coordinate-wise descend algorithms from \cite{Haghighatshoar_2018_Improved} for active UE detection. The detailed form of the algorithm is given in Algorithm \ref{al:Algorithm 1} where ML, MMV, and NNLS are short for Maximum Likelihood, Multiple Measurement Vector, and Non-Negative Least Squares, respectively. In Algorithm \ref{al:Algorithm 1}, $\bar{\mathbf{Y}} = \sqrt{\tau\rho_{p}}\boldsymbol{\Phi}\mathbf{A}\mathbf{\bar{G}}^{T} + \mathbf{\bar{W}}$ where 
$\bar{\mathbf{G}}^T = [(\mathbf{G}^1)^T, (\mathbf{G}^2)^T,..., (\mathbf{G}^N)^T]$ and $(\mathbf{G}^n)^T = [\mathbf{\bar{g}}_{[1]}^n\,\mathbf{\bar{g}}_{[2]}^n\,...\,\mathbf{\bar{g}}_{[M]}^n] \in \mathbb{C}^{K\times M}$ is the channel coefficient matrix at the first subcarrier of the $n$th subband\footnote{Note that active UE detection via Algorithm \ref{al:Algorithm 1} requires iteration. In addition, given a target $\text{P}_\text{FA}$ there is no closed-form expression for the test threshold, $\gamma'$, which requires numerical method to determine.}.
\begin{algorithm}
	\caption{Activity Detection via Coordinate-wise Descend}
	\label{al:Algorithm 1}
	\begin{algorithmic}[1]
		\item \textbf{Input}: The empirical covariance matrix $\hat{\boldsymbol{\Sigma}}_y = \frac{1}{M}\bar{\mathbf{Y}}\bar{\mathbf{Y}}^H$ of the $\tau \times MN$ matrix of samples $\bar{\mathbf{Y}}$.
		\item \textbf{Initialize}: $\boldsymbol{\Sigma} = \mathbf{I}_{\tau}, \boldsymbol{\gamma} = \mathbf{0}$.
		\item \textbf{for} $i = 1, 2, ...$ do
		\item $\;\;$ Select an index $k$ corresponding to the $k$th component of $\boldsymbol{\gamma} = [\gamma_1, \gamma_2,..., \gamma_L]^T$ (e.g., randomly).
		\item $\;\;$ \textbf{ML}: Set $d^* = \max \left\{\frac{\boldsymbol{\phi}_k^H\boldsymbol{\Sigma}^{-1}\hat{\boldsymbol{\Sigma}}_y\boldsymbol{\Sigma}^{-1}\boldsymbol{\phi}_k - \boldsymbol{\phi}_k^H\boldsymbol{\Sigma}^{-1}\boldsymbol{\phi}_k}{(\boldsymbol{\phi}_k^H\boldsymbol{\Sigma}^{-1}\boldsymbol{\phi}_k)^2}, -\gamma_k \right\}$
		\item $\;\;$ \textbf{MMV}: Set $d^* = \max\left\{\frac{\sqrt{\boldsymbol{\phi}_k^H\boldsymbol{\Sigma}^{-1}\hat{\boldsymbol{\Sigma}}_y\boldsymbol{\Sigma}^{-1}\boldsymbol{\phi}_k} - 1}{\boldsymbol{\phi}_k^H\boldsymbol{\Sigma}^{-1}\boldsymbol{\phi}_k}, -\gamma_k\right\}$
		\item $\;\;$ \textbf{NNLS}: Set $d^* = \max\left\{\frac{\boldsymbol{\phi}_k^H(\hat{\boldsymbol{\Sigma}}_y - \boldsymbol{\Sigma})\boldsymbol{\phi}_k}{\|\boldsymbol{\phi}_k\|^4}, -\gamma_k\right\}$
		\item $\;\;$ Update $\gamma_k \leftarrow \gamma_k + d^*$.
		\item $\;\;$ Update $\boldsymbol{\Sigma} \leftarrow \boldsymbol{\Sigma} + d^* \times (\boldsymbol{\phi}_k\boldsymbol{\phi}_k^H)$.
		\item \textbf{end for}
		\item \textbf{Output}: The estimated activity pattern $\boldsymbol{\gamma}$.
	\end{algorithmic}
\end{algorithm}

\subsection{LMMSE Channel Estimation with Orthogonal Pilots}
Since each subcarrier has the same pilot length in the coherence interval based pilot setting, without loss of generality we consider one subcarrier in this section so the notation for subband index is omitted. Then the received pilot signals of all active UEs are
\begin{equation}
	\label{eq:re_pilot}
	\mathbf{Y} = [\mathbf{y}_{1}\,\mathbf{y}_{2}\, ...\, \mathbf{y}_{m}] = \sqrt{\tau\rho_{p}}\boldsymbol{\Phi}\mathbf{A}\mathbf{G}^{T} + \mathbf{W},
\end{equation}
where 
$\mathbf{W}\in \mathbb{C}^{\tau \times M}$ are the noise matrix with $\sim i.i.d.\ \mathcal{CN}(0,1)$ entries. 

\subsubsection{i.i.d. Rayleigh fading}\label{sec: i.i.d. CN Channel Model}
When small-scale fading coefficients ${h_{m,k}} \sim i.i.d. \ \mathcal{CN} (0, 1)$ with respect to $m$ and $k$, $g_{m,k}$ and $g_{m',k'}$ are independent if $m \neq m'$ or $k \neq k'$. 
Based on this property, an LMMSE channel estimator for $\mathbf{G}$ is given as:
\begin{equation}
	\label{eq:E_iid}
	\mathbf{E}_\mathcal{B} = \sqrt{\tau\rho_p}\boldsymbol{\Phi}_\mathcal{B}(\mathbf{C}^{-1} + \tau\rho_p\boldsymbol{\Phi}_\mathcal{B}^H\boldsymbol{\Phi}_\mathcal{B})^{-1},
\end{equation} 
where $\mathcal{B}$ is the predicted active UE set. 
$\boldsymbol{\Phi}_\mathcal{B} = [b_1\boldsymbol{\phi}_1\, b_2\boldsymbol{\phi}_2\, ...\, b_K\boldsymbol{\phi}_K]$ where $b_k = 1$ if $k \in \mathcal{B}$ and $b_k = 0$ if $k \notin \mathcal{B}$, and $\mathbf{C} = \text{diag}\{[\beta_1, \beta_2,..., \beta_K]\}$. 
Using (\ref{eq:E_iid}), the estimated channel coefficient matrix $\hat{\mathbf{G}}\in \mathbb{C}^{K \times M}$ is $\hat{\mathbf{G}}^{T} = \mathbf{E}_\mathcal{B}^H\mathbf{Y}$.
 
\subsubsection{3GPP Channel Model based Small-Scale Fading}
When small-scale fading coefficients are generated using 3GPP channel models, channel coefficients of different antennas are correlated. Thus, for an optimal linear channel estimation, the covariance matrices $\mathbf{C}_k \in \mathbb{C}^{M\times M}, \forall k$ defined below need to be taken into account.
\begin{equation}
	\mathbf{C}_k \triangleq \text{Cov}(\mathbf{g}_k, \mathbf{g}_k),\; \mathbf{g}_k = [g_{1,k}, g_{2,k},..., g_{M,k}]^T.
\end{equation} 
By using the properties of orthogonal pilots, the LMMSE channel estimator for $\mathbf{g}_k$ is given as
\begin{equation}
	\label{eq: E_k_3GPP}
	\mathbf{E}_k = \sqrt{\tau\rho_p}\mathbf{C}_k(\tau\rho_p \mathbf{C}_k + \mathbf{I}_M)^{-1}.
\end{equation}

\begin{figure}
	\begin{center}
		\includegraphics [width=0.4\textwidth]{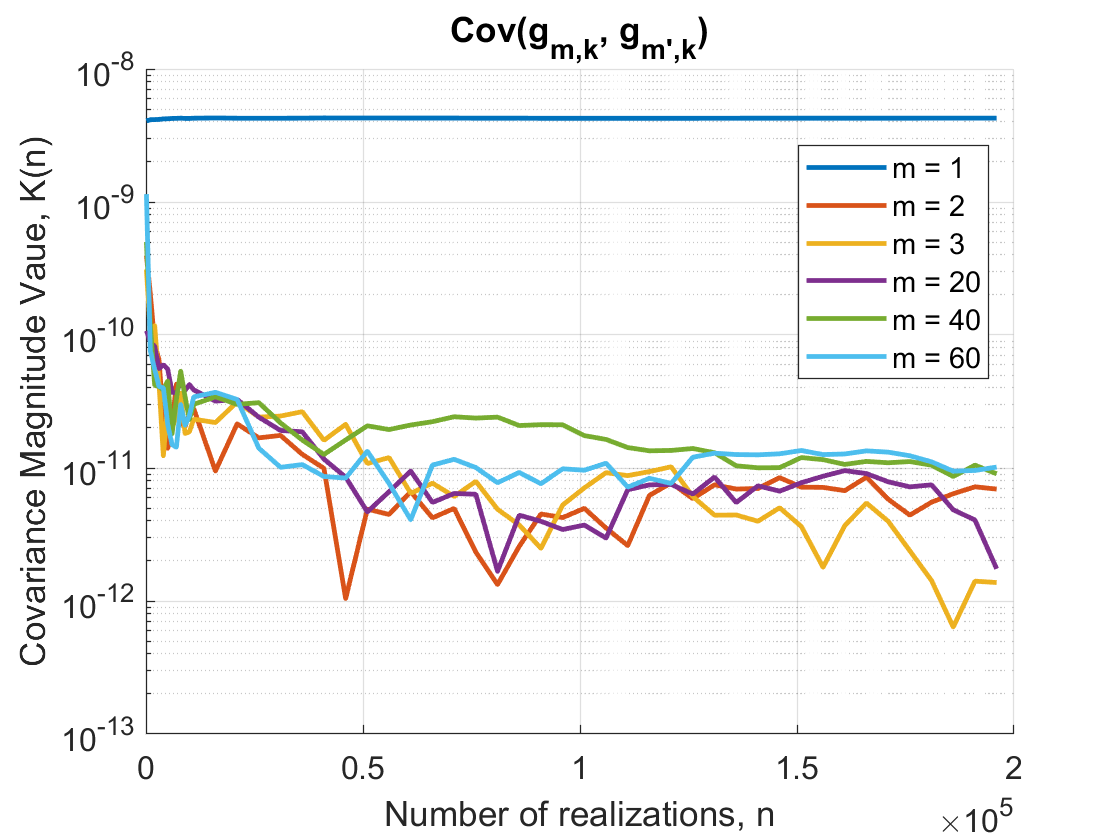}
		\caption{Channel coefficients covariance value between the first antenna and antennas $m = 2, 3, 20, 40, \text{and}\ 60.$}\label{fig:Corr_Antenna}
	\end{center}
	\vspace{-5mm}
\end{figure}

In reality, it might be unpractical to assume that matrices $\mathbf{C}_k, \forall k$ are available at the BS and hence approximations of $\mathbf{C}_k, \forall k$ are required. Based on the channel parameter settings given in Table \ref{tbl:1}, we used Nokia 3GPP channel models~\cite{3GPP_38_901} to generate channel coefficients
\begin{table}
	\renewcommand{\arraystretch}{1.4}
	\begin{center}
		\caption{3GPP Channel Model Parameters.}~\label{tbl:1}
		\fontsize{8}{8}\selectfont
		\begin{tabular}{|>{\centering\arraybackslash}m{5cm}|>{\centering\arraybackslash}m{2.5cm}|>{\centering\arraybackslash}m{0.6cm}|}\hline
			\textbf{Parameter}       &    \textbf{Value}	 \\ \hline
			Channel Name		 &    3D NR UMa NLoS 		     \\ \hline
			Carrier Frequency & 4.0 GHz \\ \hline
			BW (Bandwidth) 			     &    40 MHz			 \\ \hline
			Number of antennas ($M$)  &  128 \\ \hline 
			Array setting    &     $16 \times 8$ Planar Array  \\ \hline
			sub-carrier spacing  & 30 KHz \\ \hline
			Adjacent antenna spacing  &  half-wavelength \\ \hline
			Adjacent antenna polarization  & cross polarization\\ \hline
			Polarization angle &  $45^{\circ}$ or $-45^{\circ}$ \\ \hline
		\end{tabular}
	\end{center}
	\vspace{-3mm}
\end{table}
and found weak
channel coefficients correlation among different antennas.
Fig. \ref{fig:Corr_Antenna} shows an example of the magnitudes of the covariances between one antenna and the other antennas.
The $m = 1$ curve shows the magnitude of the channel coefficient variance. The other curves show the magnitude of channel coefficients covariances between the first antenna and the other antennas. Note that the larger $m$, the larger distances between antennas $1$ and $m$.  As we can observe, the magnitudes of the covariances are more than 500 times smaller than the magnitude of the channel coefficient variance. These large gaps indicate low correlations among antennas and thus we approximate $\mathbf{C}_k$ as
\begin{equation}
	\mathbf{C}_k \approx \hat{\mathbf{C}}_k = \text{const}_k \times \mathbf{I}_M,
\end{equation}
where $\text{const}_k = \beta_k \text{var}_\text{3GPP}$ and $\text{var}_\text{3GPP}$ is the variance of the small-scale fading coefficients generated by the 3GPP channel models. Note that $\text{const}_k$ will be equal to $\beta_k$ if small-scale fading coefficients $\sim i.i.d. \ \mathcal{CN} (0, 1)$. The procedures used in 3GPP standards for obtaining $\beta_k$ such as power head room report can also be used to obtain $\text{const}_k$. Thus we assume that the information of $\text{const}_k$ is available at the BS before the estimation of $\mathbf{G}$. Substitute $\mathbf{C}_k$ with $\hat{\mathbf{C}}_k$ in (\ref{eq: E_k_3GPP}), we obtain $\hat{\mathbf{E}}_k = \sqrt{\tau\rho_p}\hat{\mathbf{C}}_k(\tau\rho_p \hat{\mathbf{C}}_k + \mathbf{I}_M)^{-1}$. Then the estimated channel vector for the $k$th UE with $k \in \mathcal{B}$ is given as: $\hat{\mathbf{g}}_k = \hat{\mathbf{E}}_k \mathbf{Y}^T \boldsymbol{\phi}_k^*$. Note that due to the low correlation of the channel coefficients among different antennas, the active UE detection method given by (\ref{eq:T_z_k})
can be directly applied for the 3GPP generated channel coefficients.

\subsection{LMMSE Channel Estimation with Gold Sequence}
For Gold sequence, we only consider 3GPP channel model for small-scale fading. According to Fig. \ref{fig:Corr_Antenna}, the correlations among antennas are weak, thus the LMMSE channel estimation approach given in Section~\ref{sec: i.i.d. CN Channel Model} can be adopted for Gold sequences. Specifically, when using 3GPP channel models, matrix $\mathbf{C}$ in (\ref{eq:E_iid}) should be changed to $\mathbf{C} = \text{diag}\{[\text{const}_1, \text{const}_2,..., \text{const}_K]\}$.

\section{Throughput with Coherence Interval Based Pilots}\label{sec:Throughput with Coherence Interval Based Pilot Setting}
\subsection{Throughput Computation}
We adopt optimal MMSE MIMO receiver in our design and denote by $\mathbf{v}_{k}^{\text{MMSE}}$ the receiver vector for the $k$th UE. Due to the ultra-low latency requirement of URLLC, instantaneous SINR is adopted for performance evaluation and the expression for the $k$th UE is given by:
\begin{equation}
\begin{aligned}
\label{eq:SINR_k_Instans}
	&\text{SINR}_{k}^{\text{inst}}(\boldsymbol{\eta}) = \\ &\frac{\rho_{u}\eta_{k}(\mathbf{v}_{k}^{\text{MMSE}})^{H}\mathbf{g}_{k}\mathbf{g}_{k}^{H}\mathbf{v}_{k}^{\text{MMSE}}}{(\mathbf{v}_{k}^{\text{MMSE}})^{H}\left(\rho_{u}\sum_{k'\neq k, k' \in \mathcal{A}}\eta_{k'}\mathbf{g}_{k'}\mathbf{g}_{k'}^{H} + \mathbf{I}_M\right)\mathbf{v}_{k}^{\text{MMSE}}}, 
\end{aligned}
\end{equation}
where $\eta_{k}$ is the power control coefficient for the $k$th UE with the constraint $0 \le \eta_{k} \le 1$, $\rho_{u}$ is the normalized SNR of each uplink data symbol.
Using Rayleigh-Ritz Theorem, $\mathbf{v}_k^{\text{MMSE}}$ for $k \in \mathcal{B}$ can be obtained as
\begin{equation}
\label{eq:decoder_MMSE}
	\mathbf{v}_{k}^{\text{MMSE}} = \sqrt{\rho_{u}\eta_{k}}\left(\rho_{u}\sum_{k' \in \mathcal{B}}\eta_{k'}\hat{\mathbf{g}}_{k'}\hat{\mathbf{g}}_{k'}^{H} + \mathbf{I}_M\right)^{-1}\hat{\mathbf{g}}_{k}.\footnote{Based on (\ref{eq:SINR_k_Instans}) the perfect MMSE MIMO receiver have the form: $\sqrt{\rho_{u}\eta_{k}}\left(\rho_{u}\sum_{k' \in \mathcal{B}}\eta_{k'}\mathbf{g}_{k'}\mathbf{g}_{k'}^{H} + \mathbf{I}_M\right)^{-1}\mathbf{g}_{k}$. However, both perfect channel state information (CSI) and accurate statistics of the channel estimation error are not available at the BS if small-scale fading coefficients are generated according to 3GPP channel models, so we adopt the form given by (\ref{eq:decoder_MMSE}).}
\end{equation}
We use effective per-UE throughput as the performance evaluation metric. If the $k$th UE belongs to both sets $\mathcal{A}$ and $\mathcal{B}$, then its effective throughput is expressed as
\begin{equation}
	\label{eq:Throughput}
	\text{Th}_{k} = \frac{\tau_c - \tau}{\tau_c} \text{BW}\log_2 (1 + \text{SINR}_k^{\text{inst}}\times10^{-1/10}),
\end{equation}
where $10^{-1/10}$ accounts for $1$dB SINR degradation due to decoding error\footnote{For a channel with a given SINR, we can transmit at any rate $R = \log_2(1 + \text{SINR}) - \epsilon$ with $\epsilon$ an arbitrary small positive number, and achieve an arbitrary small probability of decoding error $P_{de}$ if a good, for example random, infinitely long error correcting code with optimal decoding is used. In reality a finite length code will be applied and we cannot use optimal decoding (complexity is too high). Thus, we assume that we will manage to achieve a target $P_{de}$ (e.g., $10^{-5}$) with $R = \log_2(1 + \text{SINR} \times 10^{-1/10})$.}. 
If the $k$th UE belongs to set $\mathcal{A}$ and not to set $\mathcal{B}$, this UE is mis-detected and its throughput is set as 0.

\subsection{Numerical Simulation Results}\label{sec:Numerial_Ortho}
\begin{figure}
	\begin{center}
		\includegraphics [width=0.36\textwidth]{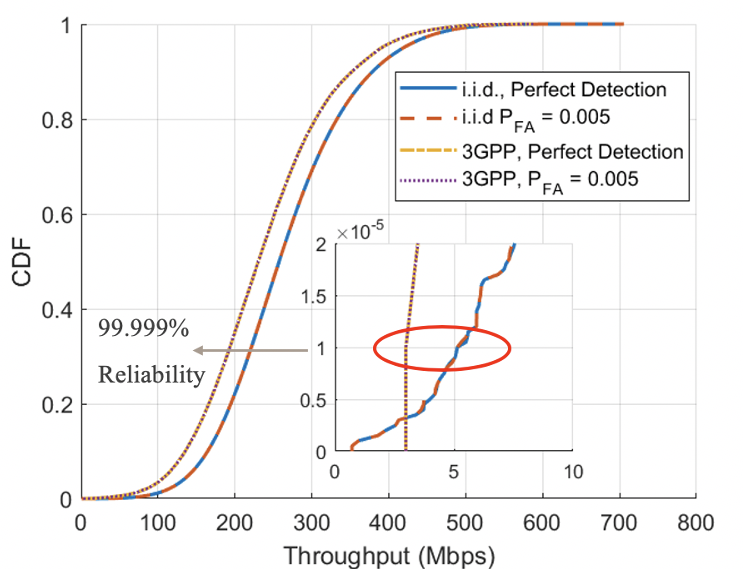}
		\vspace{-2mm}
		\caption{Per-UE throughput performance with orthogonal pilots (i.e., $\tau = 50$) under UMa NLoS scenario. Here $\text{P}_\text{FA}$ is the probability of false alarm.}\label{fig:NLoS_Orthogonal}
	\end{center}
	\vspace{-6mm}
\end{figure}

In our simulations the cell radius is set as $150$m, the number of URLLC UEs (i.e., $K$) is set as 50, and at a given moment, the average number of active UEs (i.e., $K_a$) is set as 10. To enable comparisons with the 3GPP compliant pilot setting presented later, the length $\tau_c$ of the coherence interval is set as 168 OFDM symbols, which equals the number of resource elements (RE) in a PRB. Note that in practical scenarios, $\tau_c$ depends on the mobility of environments and can be larger or less than 168. In addition, full power transmission is considered in this section.
The per-UE throughput performance with orthogonal pilots under UMa NLoS scenario is shown in Fig. \ref{fig:NLoS_Orthogonal}.
As mentioned in Section~\ref{Sec:Introduction} the URLLC requirements are 250 bytes/1ms ($2 \text{Mbps}$) with 99.999\% reliability.  Since 99.999\% reliability corresponds to the probability $10^{-5}$ in a CDF curve, we say that the URLLC requirements are satisfied if the per-UE throughput is larger than 2 Mbps at probability $10^{-5}$. As observed from Fig.~\ref{fig:NLoS_Orthogonal}, with orthogonal pilots URLLC requirements are satisfied in both i.i.d. and 3GPP small-scale fading and the probabilities of mis-detection are 0 since no UE has 0 throughput. Furthermore, the performance using NP detector is almost equal to the performance with perfect detection regime, in which we assume that the set of active UEs is known to BS and therefore $\mathcal{B} = \mathcal{A}$.

\begin{figure}
	\begin{center}
		\includegraphics [width=0.42\textwidth]{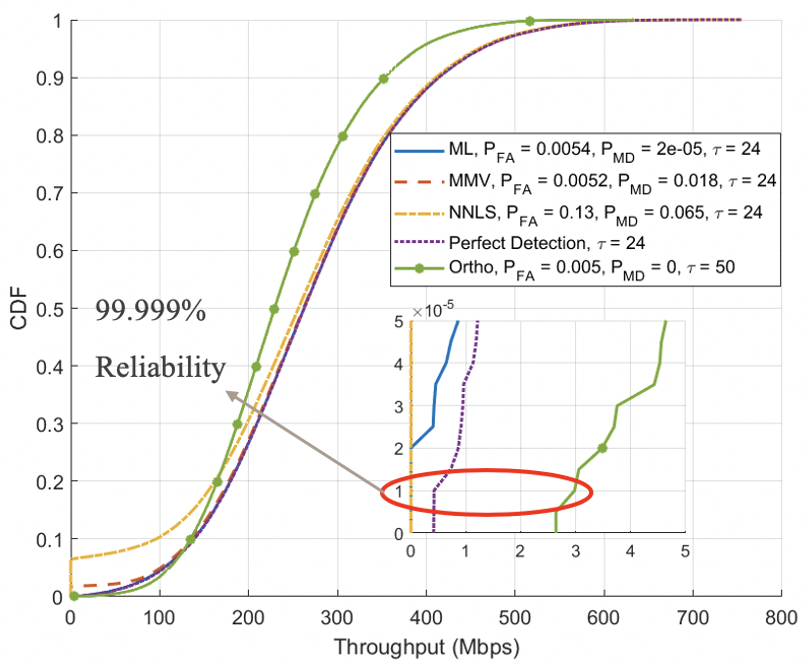}
		\vspace{-2mm}
		\caption{Per-UE throughput performance with Gold sequences ($\tau = 24$) under UMa NLoS scenario.}\label{fig:NLoS_Gold}
	\end{center}
\end{figure}
Fig. \ref{fig:NLoS_Gold} shows the per-UE throughput performance of the scenario of using Gold sequences under UMa NLoS channels. The pilot length is set as 24 which is the maximum pilot length supported by 3GPP \cite{3GPP_38_214} in a PRB, and for comparison the throughput performance of the scenario with orthogonal pilots is also included. It is observed from Fig. \ref{fig:NLoS_Gold} that even with perfect detection, the challenging URLLC requirements is not satisfied using Gold sequences. On the other hand, the performance obtained by ML detection is very close to the performance with perfect detection, but MMV and NNLS detection lead to relatively high probability of misdetection ($\text{P}_\text{MD}$) and their performance is farther from the URLLC requirements than the ML detection.
 
\section{3GPP Compliant Pilot Setting}\label{Standards pilot setting}
\begin{figure}
	\begin{center}
		\includegraphics [width=0.35\textwidth]{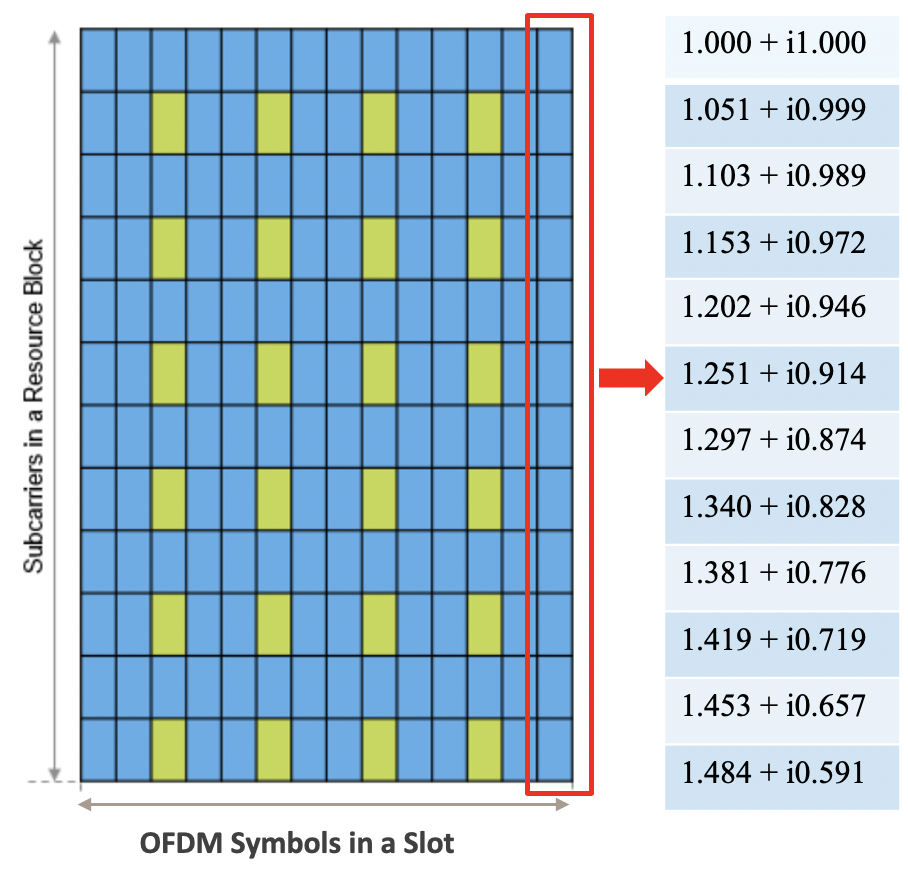}
		\vspace{-2mm}
		\caption{Pilot symbol locations and normalized channel coefficient values at different subcarriers in one PRB.}\label{fig:DM_RS_Locations}
	\end{center}
\vspace{-5mm}
\end{figure}

We consider the 3GPP compliant pilot setting in this section. An example of how pilot symbols are allocated in one PRB is shown in Fig.~\ref{fig:DM_RS_Locations}. 
The normalized channel coefficients at different subcarriers under UMa NLoS channel model are also shown. We assume the channel coefficients stay constant during the time of one slot for every subcarrier and consider the maximum pilot length (i.e., $24$).
Then the received signal at the $m$th antenna in one PRB is given by
\begin{equation}
\label{eq:y_m_tones}
	\mathbf{y}_m = \mathbf{V}_{\mathcal{A}} \tilde{\mathbf{g}}_{[m], \mathcal{A}} + \mathbf{w}_m
\end{equation}	
where the pilot matrix $\mathbf{V}_{\mathcal{A}} \in \mathbb{C}^{24 \times 6|\mathcal{A}|}$ is given as
\begin{equation}
	\label{eq:V_A}
	\mathbf{V}_{\mathcal{A}} = \left[ \begin{array}{ccccccccccccc} \mathbf{V}_{\mathcal{A}}^1 & \mathbf{0} & ... & \mathbf{0} \\ 
	\mathbf{0} & \mathbf{V}_{\mathcal{A}}^3 & ... & \mathbf{0} \\
	&  ......\\
	&  ......\\
	\mathbf{0} & \mathbf{0} & ... & \mathbf{V}_{\mathcal{A}}^{11}
	\end{array} \right],
\end{equation}
where $\mathbf{V}_{\mathcal{A}}^{s} = [\mathbf{v}_{\mathcal{A}(1)}^s \mathbf{v}_{\mathcal{A}(2)}^s ... \mathbf{v}_{\mathcal{A}(|\mathcal{A}|)}^s] \in \mathbb{C}^{4 \times |\mathcal{A}|}$. Here 
$\mathbf{v}_{\mathcal{A}(a)}^s = \boldsymbol{\phi}_{\mathcal{A}(a)}(2(s+1)- 3: 2(s+1)) \in  \mathbb{C}^{4\times 1}$, $s$ denotes the subcarrier index in a PRB and has values $s = 1, 3, ..., 11$, $\mathcal{A}(a)$ denotes the $a$th component of the active UE set $\mathcal{A}$ and $a = 1, 2, ..., |\mathcal{A}|$ where $|\mathcal{A}|$ denotes the cardinality of $\mathcal{A}$. $\boldsymbol{\phi}_{\mathcal{A}(a)}(2(s+1)- 3: 2(s+1))$ denotes the $(2(s+1)- 3)$th to the $2(s+1)$th components of the $\mathcal{A}(a)$th UE's pilot sequence. The channel vector $\tilde{\mathbf{g}}_{[m]} = [(\tilde{\mathbf{g}}_{[m], \mathcal{A}}^1)^T (\tilde{\mathbf{g}}_{[m], \mathcal{A}}^3)^T ... (\tilde{\mathbf{g}}_{[m], \mathcal{A}}^{11})^T]^T \in \mathbb{C}^{6|\mathcal{A}|\times 1}$ with $\tilde{\mathbf{g}}_{[m], \mathcal{A}}^s = [g_{m,\mathcal{A}(1)}^s, g_{m,\mathcal{A}(2)}^s,...,g_{m,\mathcal{A}(|\mathcal{A}|)}^s]^T \in \mathbb{C}^{|\mathcal{A}| \times 1}$. 
Here $g_{m, \mathcal{A}(a)}^s$ denotes the channel coefficient between the $m$th antenna of the BS and the $\mathcal{A}(a)$th UE at the $s$th subcarrier in one PRB. 

Since channel estimation is performed per PRB, we can observe from Fig.~\ref{fig:DM_RS_Locations} that at most four pilot symbols can be used for CSI acquisition per subcarrier. Under this condition, a direct application of Algorithm \ref{al:Algorithm 1} + LMMSE channel estimation in Section \ref{sec: i.i.d. CN Channel Model} leads to poor per-UE throughput performance. 
One possible way to improve the performance is to utilize the channel coefficients covariance matrix across the 12 PRB subcarriers. An estimate of this covariance matrix can be obtained with the help of the sounding reference signals (SRS) defined in the 5G NR standards. Due to space limit, we omit details on obtaining this estimation. 

\subsection{Active UE Detection}\label{Sec:Active UE Detection}
We adopt the ML approach in Algorithm \ref{al:Algorithm 1} for active UE detection but incorporate the covariance matrix among different subcarriers in a PRB. The procedures are the same as in Algorithm \ref{al:Algorithm 1} except that step 9 is changed to:
$\text{Update}\ \boldsymbol{\Sigma} \leftarrow \boldsymbol{\Sigma} + \sum_{s = 2i + 1}\mathbf{C}_k^s \boldsymbol{\phi}_k(\boldsymbol{\tilde{\phi}}_k^s)^H$ for $i = 0,1,...,5$.
Here $\mathbf{C}_k^s \triangleq \text{blkdiag}([c_k^{s,1}\mathbf{I}_4, c_{k}^{s, 3}\mathbf{I}_4,..., c_{k}^{s, 11}\mathbf{I}_4]) \in \mathbb{C}^{24 \times 24}$ where $c_k^{s, s'}$ denotes the covariance between the channel coefficients in the $s$th and the $s'$th subcarriers in one PRB for the $k$th UE with $s, s' = 1, 3, ..., 11$.
$\tilde{\boldsymbol{\phi}}_k^s \in \mathbb{C}^{24 \times 1}$ is a column vector with zero elements except the $2(s+1)-3$th to the $2(s+1)$th elements being $\boldsymbol{\phi}_k(2(s+1)-3: 2(s+1))$. 

\subsection{LMMSE Channel Estimation}
According to the form given by (\ref{eq:y_m_tones}), standard LMMSE estimation can be applied to estimate $\tilde{\mathbf{g}}_{[m]}$ and the corresponding estimator is given as
\begin{equation}
	\label{eq:E_tones}
	\tilde{\mathbf{E}}_\mathcal{B} = \sqrt{\tau\rho_p}\mathbf{V}_\mathcal{B}(\tilde{\mathbf{C}}_{\mathcal{B}}^{-1} + \tau\rho_p\mathbf{V}_\mathcal{B}^H\mathbf{V}_\mathcal{B})^{-1},
\end{equation}
where $\tilde{\mathbf{C}}_{\mathcal{B}} = [(\tilde{\mathbf{C}}_\mathcal{B}^1)^T, (\tilde{\mathbf{C}}_\mathcal{B}^3)^T,... (\tilde{\mathbf{C}}_{\mathcal{B}}^s)^T,..., (\tilde{\mathbf{C}}_\mathcal{B}^{11})^T]^T \in \mathbb{C}^{6|\mathcal{B}|\times 6|\mathcal{B}|}$, $\tilde{\mathbf{C}}_{\mathcal{B}}^{s} = [\mathbf{C}_{\mathcal{B}}^{s, 1}, \mathbf{C}_{\mathcal{B}}^{s, 3}, ..., \mathbf{C}_{\mathcal{B}}^{s, s'}, ..., \mathbf{C}_{\mathcal{B}}^{s, 11}] \in \mathbb{C}^{|\mathcal{B}|\times 6|\mathcal{B}|}$, and $\mathbf{C}_{\mathcal{B}}^{s,s'} = \text{diag}([c_{\mathcal{B}(1)}^{s, s'}, c_{\mathcal{B}(2)}^{s, s'},..., c_{\mathcal{B}(|\mathcal{B}|)}^{s, s'}])$.
Here 
$\mathcal{B}(b)$ denotes the $b$th component of the predictive UE set $\mathcal{B}$ and $b = 1,2,...,|\mathcal{B}|$.

\subsection{Numerical Simulation Results}
\begin{figure}
	\begin{center}
		\includegraphics [width=0.45\textwidth]{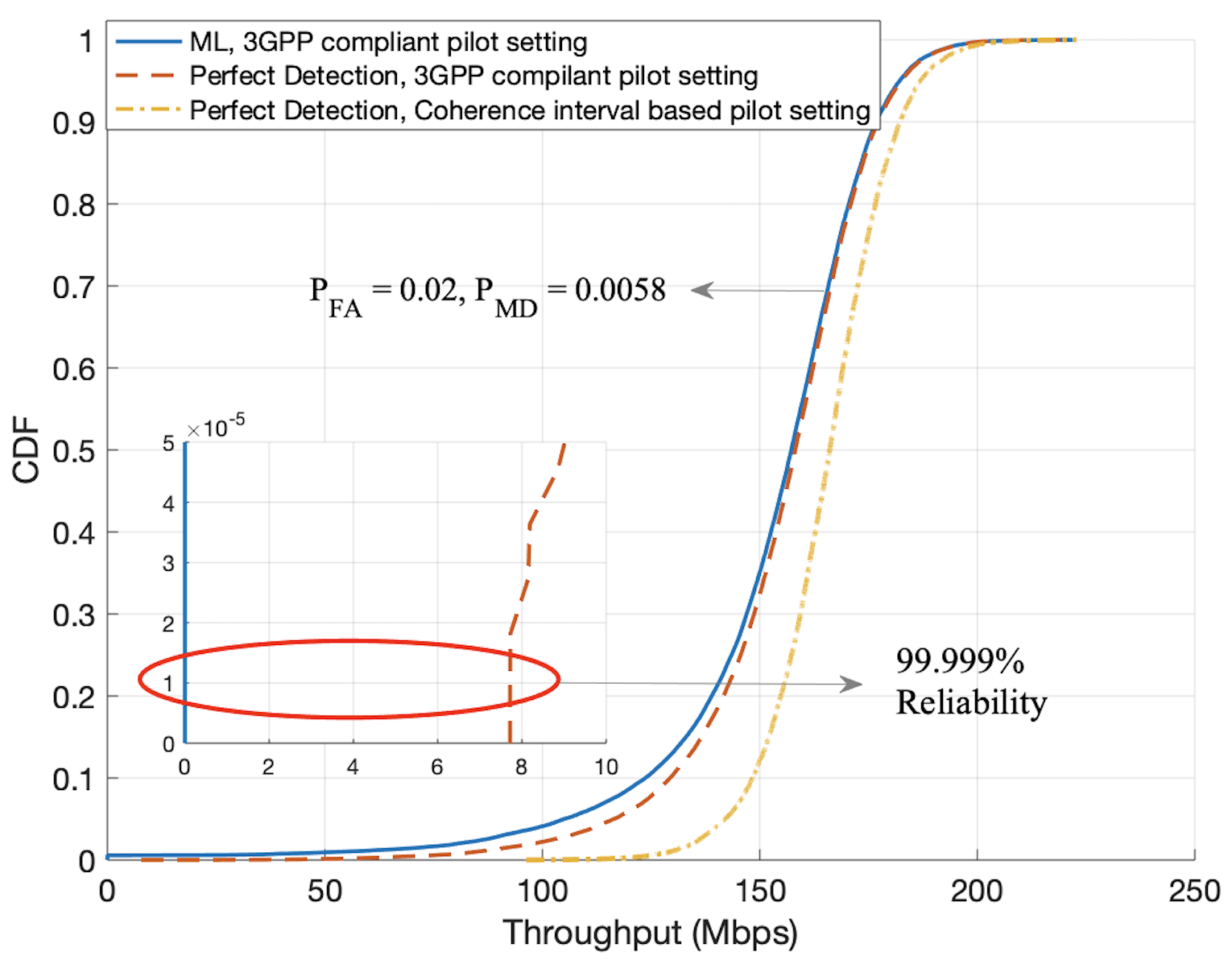}
		\vspace{-3mm}
		\caption{Per-UE throughput performance with Gold sequence under UMi NLoS scenario. Open-loop power control and dropping 5\% poor UEs are applied.}\label{fig:NLoS_Gold_Sequence_Tones_Est_Cov}
	\end{center}
\vspace{-2mm}
\end{figure}

In this section, we adopt UMi NLoS channel model with cell radius being $100$m and the remaining system settings are the same as in Table \ref{tbl:1}. We also drop $5$\% UEs with the smallest values of large-scale fading of all $\tilde{K}$ UEs. Define by $|\mathcal{R}|$ the remaining UE set after dropping poor UEs, an open-loop power control is applied which is given as
\begin{equation}
	\label{eq: open-loop}
	\eta_{\tilde{k}} = \min_{\tilde{k}}\beta_{\tilde{k}}/\beta_{\tilde{k}},\ \tilde{k} \in |\mathcal{R}|. 
\end{equation}
Fig. \ref{fig:NLoS_Gold_Sequence_Tones_Est_Cov} shows the per-UE throughput performance with Gold sequence ($\tau = 24$) under 3GPP compliant and coherence interval based pilot settings. It is observed that the performance with ML detection is very close to that with perfect detection but does not meet the URLLC reliability requirements, which indicates that these requirements can be achieved only for cells of a smaller radius, or number of service antennas $M$ should be increased. On the other hand, the performance under the coherence interval based pilot setting is significantly higher than the URLLC requirements.

\section{Summary}
We analyzed practical mMIMO supported URLLC in two pilot settings. The simulation results from the coherence interval based pilot setting show that long orthogonal pilots have significant advantage over short non-orthogonal Gold sequences and can meet URLLC requirements even without power control, i.e., with full power transmission, and with relatively large cell radius. In the 3GPP compliant pilot setting, under the assumptions that power control and the subcarrier covariance matrices are used for active UE detection and LMMSE channel estimation, meeting URLLC requirements still appears to be challenging. A reduced cell radius and/or better active UE detection algorithms are required.

\bibliographystyle{IEEEtran}
\bibliography{references}
\end{document}